\documentstyle[preprint,revtex,eqsecnum]{aps}

\tightenlines
\begin{document}
\draft

\preprint{DOE/ER/40322-170}
\preprint{U. of MD PP \#93-001}

\begin{title}
The structure of the pion and effective electroweak \\
currents in soliton models of the nucleon.
\end{title}

\author{Wojciech Broniowski \cite{ifj} and Thomas D. Cohen}
\begin{instit}
Department of Physics and Astronomy, University of Maryland \\
College Park, Maryland 20742-4111
\end{instit}

\begin{abstract}
Nonminimal substitution terms in electroweak currents are
studied in effective chiral soliton models. It is found that
the terms describing the structure of the pion lead to sizable
effects in form factors and polarizabilities of the nucleon.
\end{abstract}


\narrowtext

Much of what is known about the structure of the nucleon
and other baryons has been learned from electromagnetic and weak
probes.    For well over a decade a variety of  chiral soliton models
have been constructed  to explain the structure of the baryon.
Most of the effort in building these models has gone into the description
of the strong interaction dynamics.  It is generally assumed that the
electromagnetic and weak properties can be determined
via the study of currents obtained by gauging the
lagrangian of the model.  In this note we investigate
this assumption and show that a good description of the
strong interaction dynamics is not sufficient.  One also
needs a good description of the effective electroweak
currents, and the minimal substitution obtained by
gauging the effective theory does not necessarily
produce a good description of the effective currents.
In particular,  we point out that electroweak
properties of the nucleon ({\it e.g.} form factors, polarizabilities) in
chiral soliton models
are affected significantly by terms in the effective
lagrangian which describe the
structure of the pion.

In principle, there are an infinite number of terms
describing how a photon can couple with
pions in a nonminimal way.  Here we investigate the
effects of the $L_9$ and $L_{10}$ terms introduced
in the standard chiral expansion
Ref.~\cite{GasserL}.   We concentrate on these two
terms for the following reason --- the coefficients
associated with them can be obtained by experiments in the
meson sector, and are known with reasonable accuracy.  We
derive simple expressions relating
nonminimal-coupling contributions to nucleon observables with the pion
rms charge radius and pion polarizability
%(which can
%be expressed in terms of the radiative decay of
%of the pion \cite{Das,Terentyev,Holstein}.
We find that these nonminimal
coupling terms make sizable contributions to a number of
observable quantities.

The starting point in construction of effective
models of the nucleon structure
is the choice of an effective lagrangian with low-energy degrees of freedom.
In the large-$N_c$
approach \cite{thoft:nc,witten:nc} baryons arise as solitons of these
effective lagrangians, and semiclassical methods can be used to
describe their properties \cite{ANW83,CB86}.
Popular models based on this
idea range from purely mesonic skyrmions \cite{Skyrme6162,skyrmion:rev},
through models with quarks and mesons
(chiral quark-meson models \cite{BBC:rev},
hybrid bag models \cite{hybrid:rev}, chiral
color-dielectric models \cite{conf}), to the Nambu--Jona-Lasinio model
in the solitonic treatment \cite{NJL:sol}, which involves
quark degrees of freedom only.

In the usual approach, electroweak interactions are introduced
to an effective theory by gauging the effective lagrangian. This
minimal coupling, however, by no means guarantees
that the resulting currents correspond to currents of the underlying
QCD. In fact, the ``bosonization'' procedure, i.e. the procedure of
obtaining an effective theory from QCD, and gauging,
do not necessarily commute. Gauging QCD, and then ``bosonizing'' the
currents in general leads to a different result from first bosonizing
QCD, and then gauging the effective lagrangian. In recent years numerous
attempts \cite{derive:eff} were made to ``derive'' an effective
lagrangian from QCD. Upon expanding in powers of pion fields and their
derivatives, a comparison with the low-energy chiral perturbation
theory lagrangian \cite{GasserL} can be made. The quality of the fit in the
corresponding low-energy constants depends on the model \cite{derive:eff}.
For our present purpose the relevant terms are those describing
the structure of the pion, namely ${\cal L}_9$ and ${\cal L}_{10}$
in the notation of
Ref.~\cite{GasserL}:
\FL
\begin{eqnarray}
{\cal L}_9 &=& -\dot{\imath} L_9 \,
Tr[{\cal F}_{\mu \nu}^L D^{\mu} U D^{\nu}U^{\dagger} +
      {\cal F}_{\mu \nu}^R D^{\mu}
U^{\dagger} D^{\nu} U ] , \label{eq:Lag9}
\\
{\cal L}_{10} &=& L_{10} \,
Tr[{\cal F}_{\mu \nu}^L U {\cal F}^{\mu \nu , R} U^{\dagger}] ,
\label{eq:Lag910}
\end{eqnarray}
where \mbox{${\cal F}_{\mu \nu}^{L,R} =
{\cal V}_{\mu \nu} \mp {\cal A}_{\mu \nu}$} are
the left and right chiral field strength
tensors, and ${\cal V}_{\mu \nu}$ and ${\cal A}_{\mu \nu}$
denote the vector and axial field strength tensors:
\begin{eqnarray}
{\cal V}_{\mu \nu} = \mbox{$1\over 2$}
\tau_a (\partial_{\mu} V_{\nu}^a - \partial_{\nu} V_{\mu}^a),
\nonumber \\
{\cal A}_{\mu \nu} = \mbox{$1\over 2$}
\tau_a (\partial_{\mu} A_{\nu}^a - \partial_{\nu} A_{\mu}^a).
\label{eq:tensors}
\end{eqnarray}
The chiral field is parameterized as
\begin{equation}
U = exp(\dot{\imath} \mbox{\boldmath $\tau$}
                     \cdot \mbox{\boldmath $\phi$}) =
F_{\pi}^{-1} (\sigma + \dot{\imath} \mbox{\boldmath $\tau$} \cdot
\mbox{\boldmath $\pi$}) ,
\label{eq:chiral}
\end{equation}
where \mbox{$\sigma = F_{\pi} \cos{\phi}$},
\mbox{$\mbox{\boldmath $\pi$} = F_{\pi} \widehat{\mbox{\boldmath $\phi$}}
\sin{\phi}$}.
The constants $L_9$ and $L_{10}$ in
Eqs.~(\ref{eq:Lag9},\ref{eq:Lag910}) can be
expressed through measurable quantities \cite{GasserL,Holstein},
namely the pion isovector mean square
radius, \mbox{${\langle r^2 \rangle}^{I=1}_{\pi}$},
and the pion polarizability
\mbox{$\overline{\alpha}^{\pi}= - \overline{\beta}^{\pi}$}
\cite{GasserL,Holstein}:
\begin{eqnarray}
{L}_9 &=& \frac{F_{\pi}^2 {\langle r^2 \rangle}^{I=1}_{\pi}}{12} ,
\label{eq:L9}
\\
{L}_{10} &=& \frac{m_{\pi} F_{\pi}^2
\overline{\alpha}^{\pi}}{4 \alpha_{QED}} - L_9 =
\frac{m_{\pi} F_{\pi}^2 \alpha^{\pi}}{4 \alpha_{QED}},
\label{eq:L10}
\end{eqnarray}
where $\alpha_{QED} = 1/137$.  Although
the pion polarizability has not been measured
reliably, one can use chiral identities to
re-express this quantity in terms of the $h_A$ constant in the
decay $\pi^{+} \rightarrow \gamma \nu l^{+}$ \cite{Terentyev}.

Effective models used to describe the nucleon
have not stressed the physics in Eq.~(\ref{eq:Lag910}).
For example, the lagrangians of the
original Skyrme model \cite{Skyrme6162,ANW83}, or the chiral
quark model \cite{MCB:rev}
do not include these terms at all, while
the  constants $L_9$ and $L_{10}$ in the NJL model
\cite{NJL:sol} disagree to some extent with
the experimental values \cite{derive:eff}. For definiteness, below we
concentrate on the original Skyrme model \cite{ANW83}
and the chiral quark model \cite{BirBan8485,KRS84,EisKal}.
Other models, specifically
models with explicit vector meson degrees of freedom, will be discussed
at the end.

One remark about the coefficients in the chiral expansion
should be made before proceeding.  In the conventional
chiral perturbation theory approach in the meson sector,
one considers pionic loops, and
through renormalization ${L}_9$ and  ${L}_{10}$
acquire chiral logarithms.
In the soliton approach one treats the
effective lagrangian in the mean-field
approximation (tree level), replacing the meson
field operators by classical fields. Therefore
it is appropriate to compare low-energy constants
in soliton models with the full (renormalized)
constants from the chiral expansion.

The lagrangians used in
Refs.~\cite{Skyrme6162,ANW83,BirBan8485,KRS84,EisKal} have the generic form
\begin{eqnarray}
{\cal L}_{eff} &=& {\cal L}_{2} + ... , \nonumber \\
{\cal L}_{2} &=& \mbox{$1\over 4$} F_{\pi}^2 Tr[\partial_{\mu}
U \partial^{\mu} U^{\dagger}] ,
\label{eq:generic}
\end{eqnarray}
where the ellipses denote the Skyrme's fourth-order interaction (skyrmion),
or the coupling to quarks (chiral quark model), as well as the explicit
chiral symmetry breaking term.
The procedure of gauging replaces the derivatives in Eq.~(\ref{eq:generic})
by covariant
 derivatives:
\begin{equation}
\partial^{\mu} \rightarrow D^{\mu} = \partial^{\mu}
- \dot\imath [V^{\mu}, . ] - \dot\imath \{ A^{\mu}, . \}.
\label{eq:gauge}
\end{equation}
As a result, minimal substitution terms in the lagrangian,
${\cal L}_{min}$, appear. In the skyrmion, they
involve pionic terms arising from the quadratic and quartic terms,
\mbox{${\cal L}_{min} = {\cal L}_{min}^2 + {\cal L}_{min}^4$}.
In chiral quark model of Ref.~\cite{BirBan8485,KRS84,EisKal}
\mbox{${\cal L}_{min} = {\cal L}_{min}^2 + {\cal L}_{quark}$},
where ${\cal L}_{quark}$ is the contribution from the valence quarks.
In both types of models terms
${\cal L}_{9}$ and ${\cal L}_{10}$ are missing, and we should add them by
hand in order for the theory to describe properly the properties of the pion.
The resulting lagrangian has the form
\begin{equation}
{\cal L} = {\cal L}_{2} + {\cal L}_{min} +
{\cal L}_{9} + {\cal L}_{10} + ... ,
\label{eq:full}
\end{equation}

We first analyze the effect of the ${\cal L}_9$ term on the nucleon
electromagnetic and axial
form factors.
The minimal substitution term ${\cal L}_{min}^2$
generates the following vector and axial currents:
\FL
\begin{equation}
V_{2}^{{\nu},a} =
 {\epsilon}^{abc}{\pi}_b \partial^{\nu} {\pi}_c, \;\;
A_{2}^{{\nu},a} =
  \sigma \partial^{\nu} {\pi}^a -
   {\pi}^a\partial^{\nu} \sigma.
\label{eq:currentsmin}
\end{equation}
Additional vector and axial source currents generated by
the ${\cal L}_{9}$ term, $V_{9}$ and $A_{9}$, have the form
\FL
\begin{eqnarray}
V_{9}^{{\nu},a} &=& -4 L_{9} F_{\pi}^{-2} \partial_{\mu}
 {\epsilon}^{abc} [(\partial^{\mu} {\pi}_b) (\partial^{\nu} {\pi}_c)] ,
\nonumber \\
A_{9}^{{\nu},a} &=& -4 L_{9} F_{\pi}^{-2} \partial_{\mu}
  [(\partial^{\mu} \sigma) (\partial^{\nu} {\pi}^a) -
   (\partial^{\nu} \sigma) (\partial^{\mu} {\pi}^a)] .
\label{eq:currents}
\end{eqnarray}
Note that since these currents are total derivatives, they
do not induce additional charges. One can
also show that the magnetic moments are not changed, since
$\mbox{\boldmath $r$} \times \mbox{\boldmath $V$}_{9}^{a}$ is
also a total derivative.

In the semiclassical treatment of hedgehog soliton models, a projection
method is necessary to restore good quantum numbers of baryons.
In projection via cranking \cite{ANW83,CB86} one arrives at
semiclassical expressions, which can be obtained from field-theoretic
expressions by a substitution
\begin{equation}
\sigma \rightarrow \sigma_{h}, \;\; \pi^a \rightarrow c^{ab} \pi_{h}^b,
\;\; \partial^0 \pi^a \rightarrow - \epsilon^{abc} \lambda_b  \pi_{h}^c.
\label{eq:classical}
\end{equation}
where $\sigma^{h} = \sigma(r)$ and $\pi^a = \widehat{r}^a \pi_h(r)$ are
the usual hedgehog fields. Collective operators, $\lambda_a$ and $c^{ab}$
have the following matrix elements in the nucleon state
\begin{eqnarray}
\langle N \mid \lambda_a \mid N \rangle &=&\frac{1} {\Theta}
{\langle N \mid J_a \mid N \rangle} , \nonumber \\
\langle N \mid c^{ab} \mid N \rangle &=& - \mbox{$1\over 3$}
\langle N \mid \tau^a \sigma^b \mid N \rangle,
\label{eq:collfactors}
\end{eqnarray}
where $J$ is the spin operator, and $\Theta$
is the moment of inertia.

In the Breit frame, the form factors are defined in
the usual way:
\FL
\begin{eqnarray}
\langle {\cal N}_f(\frac{\mbox{\boldmath $q$}}{2})
      \mid J_0^{em}(0) \mid
  {\cal N}_i(- \frac{\mbox{\boldmath $q$}}{2}) \rangle &=&
  G_E(\mbox{\boldmath $q$}^2) \chi_f^{\dagger} \chi_i,  \\
\langle {\cal N}_f(\frac{\mbox{\boldmath $q$}}{2}) \mid
  \mbox{\boldmath $J$}^{em}(0) \mid
  {\cal N}_i(- \frac{\mbox{\boldmath $q$}}{2}) \rangle &=&
  \frac{G_M(\mbox{\boldmath $q$}^2)}{2 M_N}
  \chi_f^{\dagger} (\dot{\imath} \mbox{\boldmath $\sigma$} \times
     \mbox{\boldmath $q$}) \chi_i, \nonumber \\
 \\
\langle {\cal N}_f(\frac{\mbox{\boldmath $q$}}{2}) \mid
  \mbox{\boldmath $A$}^{a}(0) \mid
  {\cal N}_i(- \frac{\mbox{\boldmath $q$}}{2}) \rangle &=&
  \chi_f^{\dagger}  \frac{\tau^a}{2} \left [
  \frac{E}{M_N}
   G_A(\mbox{\boldmath $q$}^2)
  \mbox{\boldmath $\sigma$}_\perp \right.  \nonumber \\
+ \left. \left ( G_A(\mbox{\boldmath $q$}^2)
                    \right. \right. &-& \left. \left.
                      \frac{\mbox{\boldmath $q$}^2}{4 M_N^2}
 G_P(\mbox{\boldmath $q$}^2) \right )
 \mbox{\boldmath $\sigma$}_\| \right ] \chi_i, \nonumber \\
\label{eq:formfactors}
\end{eqnarray}
where $\chi_i$ and $\chi_f$ are the two-component nucleon spinors,
\mbox{$\mbox{\boldmath $\sigma$}_\| =
\widehat{\mbox{\boldmath $q$}} (\mbox{\boldmath $\sigma$} \cdot
\widehat{\mbox{\boldmath $q$}} )$},
\mbox{$\mbox{\boldmath $\sigma$}_\perp = \mbox{\boldmath $\sigma$} -
\mbox{\boldmath $\sigma$}_\|$}, and
\mbox{$E = \sqrt{M_N^2 + \mbox{\boldmath $q$}^2/4}$}.

Using Eqs.~(\ref{eq:currentsmin}-\ref{eq:formfactors}) we obtain
straightforwardly the
following relations for the additional contributions to the
nucleon isovector form factors $G_E$, $G_M$ and $G_A$,
which come from the nonminimal currents,
Eq.~(\ref{eq:currents}):
\begin{equation}
G_X^{9,I=1}(\mbox{\boldmath $q$}^2) =
 - 2 L_9 F_{\pi}^{-2} \mbox{\boldmath $q$}^2
G_X^{2,I=1}(\mbox{\boldmath $q$}^2) ,
\label{eq:ff}
\end{equation}
where \mbox{$X = E,M \;{\rm or} \; A$}, superscript
${}^2$ denotes the form factors
corresponding to the currents (\ref{eq:currentsmin}),
and superscript ${}^9$ denotes the new terms.
The relation for $G_P^{9}$ follows from PCAC, which
implies that
\begin{equation}
M_N \left ( G_A(\mbox{\boldmath $q$}^2) -
 \frac{\mbox{\boldmath $q$}^2}{4 M_N^2}
 G_P(\mbox{\boldmath $q$}^2) \right ) = \frac{
F_{\pi} m^2_{\pi} G_{\pi NN}(\mbox{\boldmath $q$})}
{( m^2_{\pi} + \mbox{\boldmath $q$}^2)}.
\label{PCAC}
\end{equation}
Since the introduction of terms ${\cal L}_9$ and ${\cal L}_{10}$
obviously does not influence $G_{\pi NN}(\mbox{\boldmath $q$}^2)$,
the following relation holds:
\begin{equation}
G_A^9(\mbox{\boldmath $q$}^2) -
 \frac{\mbox{\boldmath $q$}^2}{4 M_N^2}
 G_P^9(\mbox{\boldmath $q$}^2) = 0 ,
\label{eq:GP}
\end{equation}

In fact, Eq.~(\ref{eq:ff}) can not be trusted for
higher moments than the rms radii, since higher order effects
have not been included (this will be discussed later).
Using expressions for the moment of inertia, magnetic moments, and
$g_A$ from Refs.~(\cite{ANW83,CB86}),
and Eqs.~(\ref{eq:L9}, \ref{eq:L10}),
we arrive at the
following corrections to the mean squared radii induced by the
${\cal L}_9$ term:
\begin{eqnarray}
{\langle r^2 \rangle}_E^{9,I=1} =
\frac{\Theta_2}{\Theta_{total}} {\langle r^2 \rangle}_{\pi}^{I=1} ,
\nonumber \\
{\langle r^2 \rangle}_M^{9,I=1} =
\frac{\mu^{I=1}_2}{\mu^{I=1}_{total}} {\langle r^2 \rangle}_{\pi}^{I=1} ,
\nonumber \\
{\langle r^2 \rangle}_A^{9} =
\frac{(g_A)_2}{(g_A)_{total}} {\langle r^2 \rangle}_{\pi}^{I=1} ,
\label{eq:radii}
\end{eqnarray}
where the subscript ${}_{2}$ in
the moment of inertia $\Theta$, isovector
magnetic moment $\mu^{I=1}$, and $g_A$, denotes
contributions to mean square radii of the nucleon
from the pionic currents (\ref{eq:currentsmin}).
For example,
\begin{equation}
\Theta_2 = \mbox{$2\over 3$} \int d^3 x \,\pi_h^2 .
\label{eq:momin}
\end{equation}

In the Skyrme model of Ref.~\cite{ANW83,Adkins84} we find
${\Theta_2}/{\Theta_{total}} =
{\mu^{I=1}_2}/{\mu^{I=1}_{total}} \simeq
0.4 - 0.6$, ${(g_A)_2}/{(g_A)_{total}} \simeq 0.5$. In the chiral
quark model one finds \cite{CB86}
${\Theta_2}/{\Theta_{total}}
\simeq {\mu^{I=1}_2}/{\mu^{I=1}_{total}} \simeq 0.6$,
${(g_A)_2}/{(g_A)_{total}} \simeq 0.5$.
Since in experiment
${\langle r^2 \rangle}_{\pi} = 0.44 fm^2$, Eqs.~(\ref{eq:radii})
lead to  $\sim 0.2 fm^2$ contributions to the mean squared radii.

Experimental values for the nucleon radii are \cite{tables}:
${\langle r^2 \rangle}_E^{I=1} = 0.82 fm^2$,
${\langle r^2 \rangle}_M^{I=1} = 0.73 fm^2$,
${\langle r^2 \rangle}_A = 0.65 fm^2$.
Therefore the corrections from Eqs.~(\ref{eq:radii}) are
substantial: $\sim 25\%$, $\sim 30\%$ and $\sim 30\%$ for the
three radii squared, respectively. For a quantity which
involves cancellations between isoscalar and isovector
properties, namely the electric radius of the neutron, the
${\cal L}_9$ effect is of the order of $100\%$ !
This shows that predictions for the neutron form factor
are extremely sensitive to nonminimal coupling terms.

Now we turn to the nucleon electromagnetic polarizabilities.
This issue has been discussed in detail in Ref.~\cite{pol:ours}.
We assume that the electric and magnetic fields are constant.
Using our semiclassical methods we can
easily identify pieces in the lagrangian which are quadratic in the
$E$ and $B$ fields. The ${\cal L}_{9}$ and ${\cal L}_{10}$
terms leads to the following
semiclassical expression:
\FL
\begin{eqnarray}
\int d^3 x \,{\cal L}_9 &=&
2 \; \alpha_{QED} L_9 F_{\pi}^{-2}(E^2 - B^2)
\int d^3 x \,\mbox{$2\over 3$}\pi_h^2, \nonumber \\
\int d^3 x \,{\cal L}_{10} &=& 2 \; \alpha_{QED} L_{10} F_{\pi}^{-2}
(E^2 - B^2)   \int d^3 x \, \mbox{$2\over 3$}\pi_h^2 .
\label{eq:L910int}
\end{eqnarray}

The coefficient of $E^2$ ($B^2$) in the lagrangian has the interpretation
of twice the electric (magnetic) polarizability
\cite{pol:ours,Chemtob87}. Recognizing
the moment of inertia $\Theta_2$ in the integrals
in Eqs.~(\ref{eq:L910int}), and using Eq.~(\ref{eq:L9}, \ref{eq:L10}),
we obtain the following expression
the contributions due to the ${\cal L}_{9}$ and
${\cal L}_{10}$ terms to the electric ($\alpha_N^{\pi}$) and
magnetic ($\beta_N^{\pi}$) polarizabilities on the nucleon:
\begin{equation}
\alpha_N^{\pi} = -\beta_N^{\pi} = m_{\pi} \Theta_{2} {\overline\alpha^{\pi}} .
\label{eq:alphaNpi}
\end{equation}

Note the opposite signs of the electric
and magnetic polarizabilities in
(\ref{eq:alphaNpi}), reflecting the fact that
$\overline{\alpha^{\pi}} = - \overline{\beta^{\pi}}$.
Numerical values for $m_{\pi} \Theta_{2}$ are of the order
$\sim 0.5 - 0.7$ in the discussed models.
The value of ${\overline\alpha^{\pi}}$ can be determined
experimentally, however existing
experimental data \cite{pionpolexp} do not seem
reliable, and are in
contradiction with a low-energy theorem
\cite{Das,Terentyev,Holstein}, which gives
${\overline\alpha^{\pi}} = 2.8 \times 10^{-4} fm^3$. With this value
we get
\begin{equation}
\alpha_N^{\pi} = -\beta_N^{\pi} \sim 1.3 \times 10^{-4} fm^3 ,
\label{eq:alphaNpinum}
\end{equation}
which is a few times smaller compared to the minimal substitution terms,
\cite{pol:ours} but
non-negligible, especially for the magnetic case, where large
cancellations are expected. If experimental
numbers for $\overline\alpha^{\pi}$ were used
\cite{pionpolexp}, then a two-three times larger result follows.

The physical interpretation of the results in Eqs.~(\ref{eq:radii},
\ref{eq:alphaNpi}) is clear. Since the pion has
electromagnetic structure, it gives rise to an additional
contribution to the nucleon electromagnetic properties in
models which have a pion cloud. As an example, let us
consider the electric isovector form factor.
Suppose $\rho_{cloud}(\mbox{\boldmath $r$})$ describes the
distribution of the classical pion cloud around the nucleon,
and $\rho_{\pi}(\mbox{\boldmath $r$})$ describes
the distribution of charge in the pion.
Folding of these two distributions results in the charge form factor
of the nucleon of the form $\tilde\rho_{cloud}(\mbox{\boldmath $q$})
\tilde\rho_{\pi}(\mbox{\boldmath $q$})$.
In fact, Eqs.~(\ref{eq:ff}) can be
viewed as a term in the expansion of this product of form factors.
 From a different point of view, namely from a description
in terms of a field theory at the  hadronic level,
the electromagnetic current can couple to the nucleon via pionic loop.
The coupling to the pion involves a form factor, and this obviously
modifies the properties of the nucleon.

Up to this point we have not discussed how the  the
${\cal L}_{9}$ and ${\cal L}_{10}$ terms included
in our analysis behave in terms of $N_c$ counting or chiral counting.
It is easy to verify that since ${\langle r^2 \rangle}_{\pi}^{I=1}$,
$\overline\alpha^{\pi}$ and $m_{\pi}$ are of the order $1$ in $N_c$-counting,
and $\Theta$ is of the order $N_c$, the
additional contributions to radii, Eq.~(\ref{eq:radii}), are of the order
$1$, and for polarizabilities, Eq.~(\ref{eq:alphaNpi}) of the order $N_c$.
These are the same orders as for the contributions
resulting from the minimal coupling, hence the ${\cal L}_{9}$ and ${\cal
L}_{10}$
effect occur at the leading $N_c$ level.

In chiral counting these terms are assigned
power $4$ in momenta, whereas the minimal coupling terms
have the chiral order $2$. The difference in the application
of chiral lagrangians to the nucleons rather than mesons is
that the terms with more gradients are not necessarily suppressed.
The gradients in a soliton are of the order of $1 GeV$, which is the same
order as the chiral scale ($\sim 4 \pi F_{\pi}$, or $m_{\rho}$).
In our opinion there is no reason to expect why terms such as
${\cal L}_{9}$ and ${\cal L}_{10}$ should be suppressed in soliton models.

A more worrying issue is that one the ${\cal L}_{9
}$ and ${\cal L}_{10}$  are just two of an  infinite class of terms
which can modify the electroweak currents. For example, one could add
a term as ${\cal L}_{9}$  multiplied by
$Tr[\partial^{\sigma} U \partial_{\sigma} U^{\dagger}]^n$,
(where n is an arbitrary positive integer), which
has the same $N_c$-counting as ${\cal L}_{9}$.   Although
such a term is a higher dimensional operator and as
such is suppressed in chiral counting in the meson sector,
as mentioned above it is {\it not} suppressed  in a soliton.
In principle, one  can fix the
coefficients for these terms by from experimental studies
of interactions involving virtual photons and several pions.
In practice, however, we will never be able to determine
the phenomenological low-energy
constants for any but a few of such terms,
and thus predictive power is lost.    {\it A priori}
there is no reason to believe that the effect of these
higher dimensional operators on the effective currents
should be small.

In two classes of chiral models the physics
of ${\cal L}_{9}$ and ${\cal L}_{10}$
terms is included. One is the class of
models with explicit vector meson degrees of
freedom \cite{vector:all}. Since ${\cal L}_{9}$ basically describes the
physics of vector meson
dominance, such models are expected to give reasonable value of the
constant $L_9$. The other class are the NJL models \cite{NJL:sol}.
Upon expanding
one-quark loop in presence of external sources, these models generate
low-energy constants. Certainly, the quality of the fit depends on the
details of the model (cut-off scheme, etc.). In the case of models
which generate terms  ${\cal L}_{9}$ and ${\cal L}_{10}$, the corrections
of Eqs.~(\ref{eq:radii},\ref{eq:alphaNpi}) involve not the phenomenological
constants $L_9$ and $L_{10}$, but the differences of phenomenological
and model-predicted values

To summarize, the effective electromagnetic and weak currents
in chiral soliton models need not be the ones
obtained by minimal substition. We have demonstrated
important effects of the
nonminimal terms in the effective lagrangian on electroweak
properties of the nucleon, associated with the
${\cal L}_{9}$ and ${\cal L}_{10}$ terms.
Effects  of the order of $20 - 30 \%$ are found
for various nucleon mean radii squared in the original
Skyrme model, or simplest version of the
simple chiral quark model.    These corrections
are clearly substantial.  In addition to
the effects which we know how to estimate,
however, there are infinite number of additional terms,
with unknown constants,  which can alter the
effective currents.  The contributions for these
additional terms are not {\it a priori} small, and
they enter at leading order in the $1/N_c$ expansion.

We should also add that the effects of the
${\cal L}_{9}$ and ${\cal L}_{10}$ terms are large in models
in which the pion dominates. In models where the pion is not the
key dynamical factor, e.g. the models with confinement \cite{conf},
these effects are weaker, since the ratios in Eqs. (\ref{eq:radii})
are significantly less than $1$.

Support of the the National Science Foundation (Presidential Young
Investigator grant), and of the U.S. Department of Energy is gratefully
acknowledged. We thank Manoj Banerjee for many helpful discussions.
One of us (WB) acknowledges a partial support of
the Polish State Committee for Scientific Research (grants 2.0204.91.01
and 2.0091.91.01).

\end{document}